\documentclass[reprint, aps, prl, notitlepage]{revtex4-2}

\usepackage{amsmath, amsfonts, amssymb, amsthm, mathtools}
\usepackage[utf8]{inputenc}
\usepackage[T1]{fontenc}
\usepackage{hyperref}
\usepackage{graphicx}
\usepackage{caption}
\usepackage{subcaption} 
\usepackage{lipsum}
\usepackage{xcolor}
\usepackage{empheq}

\usepackage{pgffor}
\usepackage[]{pdfpages}

\makeatletter
\AtBeginDocument{\let\LS@rot\@undefined}
\makeatother

\def\supplementfilename{supplementDEFv4.pdf}
\pdfximage{\supplementfilename}
\def\numbersupplementpages{\the\pdflastximagepages}
\newif\ifsupp
\supptrue 

\theoremstyle{remark}

\theoremstyle{remark}

\theoremstyle{theorem}

\theoremstyle{theorem}

\newcommand{\LT}[1]{\widetilde{#1}}
\newcommand{\FT}[1]{\widehat{#1}}

\begin{document}
	
\title{Optimizing Leapover Lengths of Lévy Flights with Resetting}
\author{Mattia Radice}
\email[Corresponding author: ]{mradice@pks.mpg.de}
\affiliation{Max Planck Institute for the Physics of Complex Systems, 01187 Dresden, Germany}
\author{Giampaolo Cristadoro}
\affiliation{Dipartimento di Matematica e Applicazioni, Università degli Studi Milano-Bicocca, Milan, Italy}
\begin{abstract}
We consider a one-dimensional search process under stochastic resetting conditions. A target is located at $b\geq0$ and a searcher, starting from the origin, performs a discrete-time random walk with independent jumps drawn from a heavy-tailed distribution.  Before each jump, there is a given probability $r$ of restarting the walk from the initial position. The efficiency of  a “myopic search”\textemdash in which the search stops upon crossing the target for the first time\textemdash is usually characterized in terms of the first-passage time $\tau$. On the other hand, great relevance is encapsulated by the leapover length $l = x_{\tau} - b$, which measures how far from the target the search ends. For symmetric heavy-tailed jump distributions, in the absence of resetting the average leapover is always infinite. Here we show instead that resetting induces a finite average leapover $\ell_b(r)$ if the mean jump length is finite. We compute exactly $\ell_b(r)$ and determine the condition under which resetting allows for nontrivial optimization, i.e., for the existence of $r^*$ such that $\ell_b(r^*)$ is minimal and smaller than the average leapover of the single jump.
\end{abstract}

\maketitle

\begin{figure*}
	\centering
	\includegraphics[width=.8\linewidth]{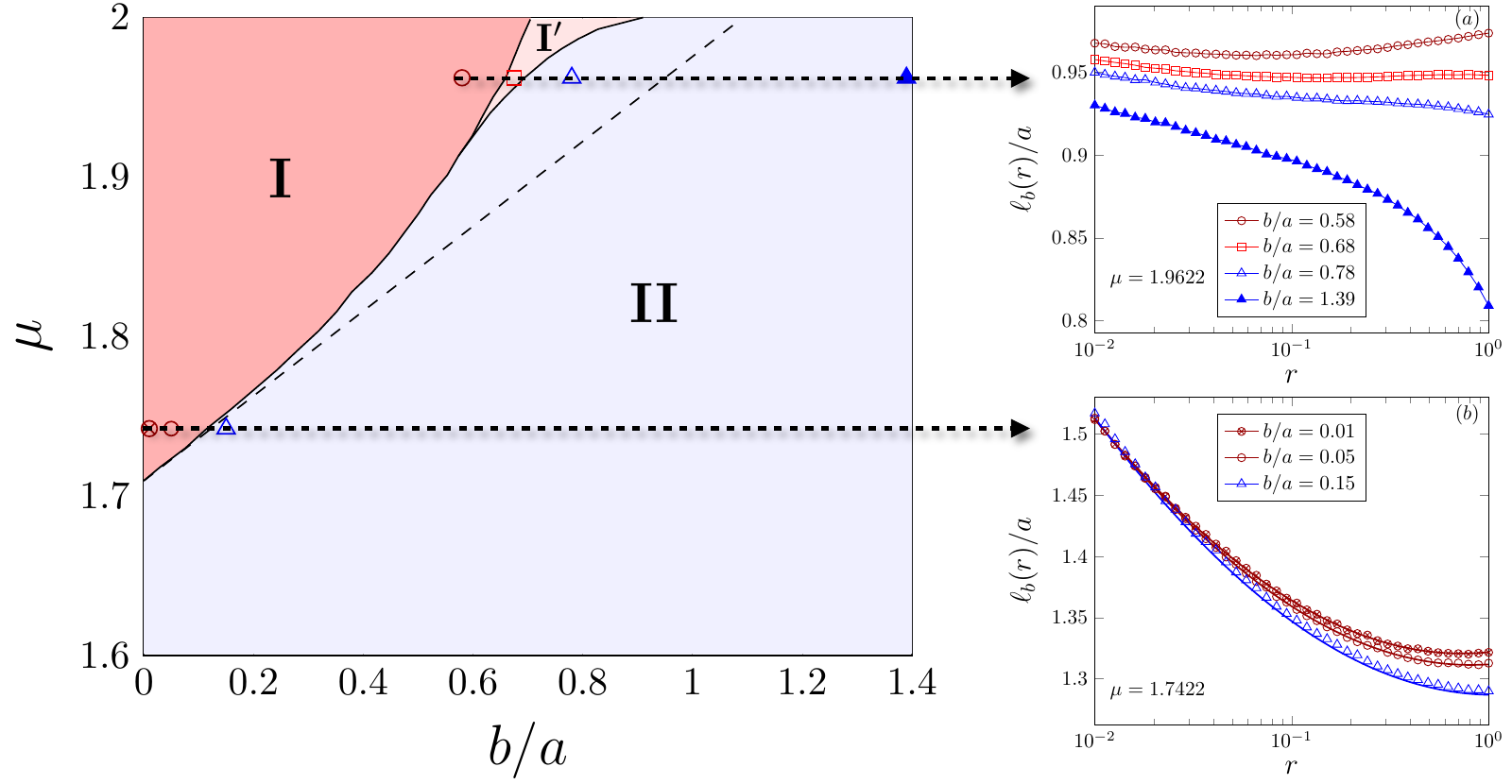}
	\caption{Phase diagram for the case of Lévy stable jump processes, with $\hat{\lambda}(k)=e^{-|ak|^\mu}$. Region (I) is the phase where a nontrivial optimal resetting probability $0<r^*<1$ is guaranteed to exists: in this regime  the optimal average leapover  $\ell_b(r^*)$ is smaller than that obtained by a single jump. The boundary of this region is numerically computed from the exact condition given in Eq. \eqref{eq:Cond}. The dashed straight line corresponds to the analytical linear approximation of the boundary for small $b$, i.e., for targets close to the origin, given by Eq. \eqref{eq:mu_c_linear}.
	In Region (I$'$) the sufficient condition Eq. \eqref{eq:Cond} is not satisfied but a nontrivial optimization probability is still observed, together with a local maximum for  $\ell_b(r)$  (see panel (a) for details). This region extends up to $b/a\approx0.91$. Region (II) is the phase where  the average leapover  $\ell_b(r)$ assumes its minimum at the “trivial” value $r=1$. The data displayed in panels (a) and (b), obtained by numerical simulations, show examples of transitions from one region to another as $b$ increases, with $\mu$ kept fixed. In (b) we also show a comparison with the small-$b$ approximation of $\ell_b(r)$ given by Eq. \eqref{eq:ell_b_approx}.}
	\label{fig:muc_b}
\end{figure*}

In various situations found in nature on both macroscopic and microscopic scales, the analysis of the strategies that a searcher implements to reach a target proves to be of extreme interest \cite{BenLovMor-2011}. Search strategies are relevant, for example, in the study of protein-DNA interactions \cite{BenLovMor-2011}, diffusion-controlled reactions \cite{Kotomin1996}, and the movement of animals or microorganisms searching for food \cite{Bell,Klafter1990}. The fundamental question common to all these different settings concerns the search efficiency and the existence of an optimal protocol to perform the task. The highly complex underlying dynamics are usually well described by random walk models, and thus the problem reduces to the study of optimization strategies for such stochastic processes.

In many contexts a central role is played by jump processes, which are defined as a sequence of independent jumps of random lengths \cite{LevBenVoi-2021,KliVoiBen-2022,KliVoiBen-2024,VezBarBur-2024}. For these processes a “myopic search” is typically implemented, where the search stops once the walker jumps over the target for the first time \cite{KusMajSabSch-2014,PalBlaLom-2019,PadCheDyb-2019,PadCheDyb-2020}. The final distance to the target, known in the literature as \emph{leapover} length \cite{Kor-Lom-al}, is therefore an aspect that should not be underestimated in evaluating the efficiency of a myopic search \cite{PalCheMet-2014}. This is especially relevant for Lévy flights, a well-known class of random walks with independent jumps drawn from a common probability density function (PDF) decaying for large $\eta$ as $\lambda(\eta)\propto\eta^{-1-\mu}$, with $0<\mu<2$. Indeed, it is known that for Lévy flights with symmetric jumps the PDF of the leapover length decays as $\wp(l)\propto l^{-1-\mu/2}$, leading to the seemingly paradoxical result that  the average leapover diverges also in the range $1<\mu<2$, where the mean jump length is finite \cite{Kor-Lom-al,GodMajSch-2016}. This feature is far from a subtlety, as Lévy flights are ubiquitous in nature and have found applications ranging from animal foraging \cite{VisAfaBul-1996,Vis-2010} to human behavior \cite{BroHufGei-2006}. Hence $l$ is important to quantify, e.g., how far an animal gets from the resources it is looking for, if it crosses them in a \emph{blind} phase of the search \cite{BenLovMor-2011}, or how far an infectious disease spreads after the first crossing  of a border \cite{BroHufGei-2006}. We also observe that leapover lengths are not only crucial in the context of search efficiency, but also play an important role in extreme value theory, where they can be interpreted as record increments \cite{GodMajSch-2016}, and in renewal theory, where they admit an interpretation as waiting times in the presence of aging \cite{Barkai-2003}.

In recent years, one of the most studied strategies for search optimization is \emph{stochastic resetting}, which consists in restarting the search after random time intervals until the target is reached or crossed \cite{EvaMaj-2011,EvaMajSch-2020}. The main advantage of this strategy is the ability to control the search time $\tau$ through an external parameter (typically, the reset rate \cite{EvaMaj-2011,EvaMajSch-2020}, but also different mechanisms have been studied \cite{PalKunEva-2016,PalReu-2017}), which can be optimally chosen to minimize the average capture time of the target. This has been proven effective in a variety of areas, ranging from biology \cite{ReuUrbKla-2014,RolLisSanGri-2016} to algorithm optimization \cite{MonZec-2002}, and has therefore generated increasing interest, both theoretical and experimental \cite{KusMajSabSch-2014,Reu-2016,CheSok-2018,TalPalSek-2020,BesBovPet-2020,BonPal-2021,BonPalReu-2022,WanCheKan-2021,VinCheWan-2022,WanCheKan-2022}.

In this Letter, we show that stochastic resetting is an effective strategy to control the leapover length of Lévy flights. We obtain exactly the leapover distribution and show that for $ \mu>1$ resetting induces a finite average leapover, which we compute analytically (see Eq. (S2) in Supplemental  Material \cite{supp}). We then determine a sufficient condition for the existence of a \emph{nontrivial} optimization strategy, i.e., such that the minimal average leapover under resetting is strictly smaller than that obtained from a single jump, see Eq. \eqref{eq:Cond}. This condition, which strikingly involves properties of the reset-free random walk after $1$ and $2$ jumps only, makes it possible to identify regions in the $(\mu, b)$ plane where such a nontrivial optimization is possible, see Fig. \ref{fig:muc_b} for a paradigmatic example.

To demonstrate these results, we consider a one-dimensional random walk $x_n$ that for $n\geq1$ evolves according to
\begin{equation}\label{eq:wlk}
	x_{n}=\begin{cases}
		x_{n-1}+\eta_{n}&\text{with probability }1-r\\
		\eta_{n} &\text{with probability }r,
	\end{cases}
\end{equation}
where $0\leq r\leq1$ and $\eta_n$ are independent continuous random variables with common PDF $\lambda(\eta)$. The initial position is $x_0=0$. A target is placed at $b\geq0$ and the evolution goes on until the \emph{first-passage time} in $(b,\infty)$, denoted as $\tau$. Consequently, $l=x_\tau-b$ is the leapover length. We call $f_r(x,n;b)$ the joint PDF of $x_\tau$ and $\tau$. The leapover PDF is thus given by $\wp_r(l;b)=\sum_{n\geq1}f_r(l+b,n;b)$. Let us also introduce $q_0(x,n;b)$, which is the PDF of the position $x_n$ for walks that, in the absence of resetting, never cross $b$ in $n$ steps. We can then set up a first-renewal equation for $f_r(x,n;b)$ \cite{KusMajSabSch-2014}:
\begin{align}
	f_r(x,n;b)=&r\sum_{m=0}^{n-1}(1-r)^{m}Q_0(m;b)f_r(x,n-m;b)\nonumber\\
	&+(1-r)^{n}f_0(x,n;b),\label{eq:f_r_xn}
\end{align}
 where $Q_0(n;b)=\int_{-\infty}^{b}q_0(x,n;b)dx$ is the \emph{survival probability} in $n$ steps. By defining the double transform $\mathcal{G}(k,z)=\sum_{n\geq0}z^n\int_{-\infty}^{+\infty}e^{ikx}g(x,n)dx$, we obtain from Eq. \eqref{eq:f_r_xn} a relation between $\mathcal{F}_r(k,z;b)$ and $\mathcal{Q}_0(k,z;b)$, the transforms of $f_r(x,n;b)$ and $q_0(x,n;b)$ respectively. The Fourier transform of the leapover PDF is then obtained from $\FT{\wp}_r(k;b)=e^{-ikb}\mathcal{F}_r(k,1;b)$, and we arrive at 
\begin{equation}\label{eq:F_cal}
	\FT{\wp}_r(k;b)=e^{-ikb}\frac{1-[1-(1-r)\hat{\lambda}(k)]\mathcal{Q}_0(k,1-r;b)}{1-r\mathcal{Q}_0(0,1-r;b)},
\end{equation}
where $\hat{\lambda}(k)=\int_{-\infty}^{+\infty}e^{ik\eta}\lambda(\eta)d\eta$ is the characteristic function of the jumps. From this exact expression one can obtain, e.g., the large-$l$ behavior of the leapover distribution and the average leapover length $\ell_b(r)$. We point out that the only unknown on the right-hand side of Eq.\eqref{eq:F_cal} is $\mathcal{Q}_0(k,z;b)$. By using a result of Spitzer \cite{Spi-1956}, we compute exactly its Laplace transform with respect to $b$, which reads
\begin{equation}\label{eq:Q_0_b>0}
	\int_{0}^{\infty}e^{-wb}\mathcal{Q}_0(k,z;b)db=\mathcal{Q}_0(k,z)\frac{e^{\phi_+(k+iw,z)}}{w},
\end{equation}
where $\mathcal{Q}_0(k,z)$ is a short-hand notation for $\mathcal{Q}_0(k,z;b=0)$ and is known in the literature \cite{Zum-Kla}, whereas $\phi_+(\xi,z)$ is defined as
\begin{equation}
	\phi_+(\xi,z)=\frac{1}{2\pi i}\int_{-\infty}^{+\infty}\frac{ds}{\xi-s}\ln(1-z\hat{\lambda}(s)).
\end{equation}
It can be shown that $\phi_+(\xi,z)$ is analytic for $\Im(\xi)>0$. We can thus compute the small-$k$ expansion of the rhs in Eq. \eqref{eq:Q_0_b>0}, and then transform it back to obtain the small-$k$ expansion of $\mathcal{Q}_0(k,z;b)$. 
Finally, we can derive the expansions for both the real and imaginary parts of $\FT{\wp}_r(k;b)$.

We consider the case of symmetric jump distributions that for large $|\eta|$ decay as $\hat{\lambda}(\eta)\propto|\eta|^{-1-\mu}$, with $0<\mu<2$. Consequently, $\hat{\lambda}(k)\sim1-|ak|^\mu$ for small $k$, where $a$ is a scale parameter. For $r>0$, we obtain $\operatorname{Re}(\FT{\wp}_r(k;b))\sim1-A_r(b)|ak|^\mu$, where $A_r(b)$ is written in terms of the generating function $\LT{Q}_0(z;b)=\sum_{n\geq0}z^nQ_0(n;b)$ \cite{supp}. Hence $\wp_r(l;b)\sim A_r(b)\lambda(l)$ for large $l$. It follows that, as announced, for $\mu>1$ the average leapover length $\ell_b(r)$ is finite and can be expressed as a well-defined function of $r$. Indeed, for small $k$ and $\mu>1$ we obtain $\operatorname{Im}(\FT{\wp}_r(k;b))\sim k\ell_b(r)$, where $\ell_b(r)$ can be determined exactly \cite{supp}. For $r=1$, this function is equal to the average leapover of a single jump. On the other hand, for $r\to0$ we must recover the behavior of the reset-free walk, yielding a diverging $\ell_b(r)$ for $\mu<2$. Hence, in the range $1<\mu<2$, since $\ell_b(r)$ diverges for $r\to0$ and is finite for $r=1$, a sufficient condition for the existence of a probability $0<r^*<1$ that minimizes $\ell_b(r)$ is that the derivative $\ell'_b(r)$ be positive as $r\to1$. By using our exact result, we find that this condition can be recast as
\begin{equation}\label{eq:Cond}
	\frac{\langle|x_2|\rangle}{\langle|x_1|\rangle}< \Psi_\mu(b),
\end{equation}
where both sides of this inequality depend only on the first two steps of the reset-free walk, thus it is easy to evaluate numerically. Here $\langle|x_n|\rangle$ is the absolute first moment of the walk after $n$ steps and $\Psi_\mu(b)$ has an exact expression (see Eq. (S4) in \cite{supp}) that involves $\lambda(\eta)$ and the survival probability in one and two steps. Interestingly, for $b=0$ the rhs is a constant, $\Psi_\mu(0)=\frac32$, independently of the jump distribution, as long as it is symmetric and continuous. We stress that, as both sides have an implicit dependence on $\mu$, Eq. \eqref{eq:Cond} allows one to determine the existence of a region in the $(\mu,b)$ plane where $\ell_b(r)$ has a global minimum at some $0<r^*<1$. In Fig. \ref{fig:muc_b} we draw the phase diagram for the paradigmatic example of the family of Lévy stable processes, whose jumps have the characteristic function
$\hat{\lambda}(k)=e^{-|ak|^\mu}$. 

\begin{figure*}
	\centering
	\includegraphics[width=.9\linewidth]{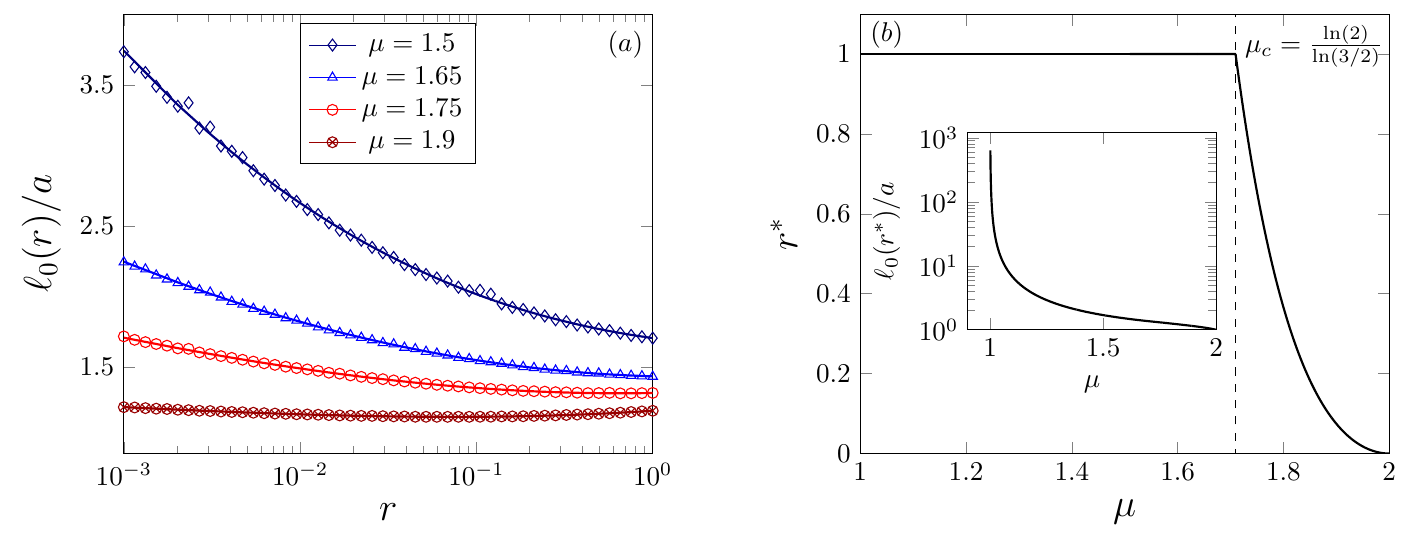}
\caption{(a) Average leapover distance and (b) optimal resetting probability for Lévy stable processes, case $b=0$. In (a) the results of numerical simulations, displayed in function of the resetting probability, are compared with the theoretical curves (see Eq. (S55) in \cite{supp}), showing excellent agreement. In (b) we present $r^*$ and (inset) the corresponding $\ell_0(r^*)$ versus $\mu$. The dashed vertical line represents the critical value $\mu_c$. The values of $r^*$ are obtained by evaluating numerically the minimum of $\ell_0(r)$.}
\label{fig:r_min}
\end{figure*}

We now consider in detail the case $b=0$. The case of a target at the origin is particularly significant because, thanks to the Sparre-Andersen universality \cite{Sparre}, we are able to provide a compact expression for the average leapover length. Indeed, the exact formula for $\ell_b(r)$ assumes a simple form when $b=0$:
\begin{subequations}
	\begin{align}
		\ell_0(r)&=\frac{\sqrt{r}}{\pi(1-\sqrt{r})}\int_{0}^{\infty}\frac{ds}{s^2}\ln\left(\frac{1-(1-r)\hat{\lambda}(s)}{r}\right)\label{eq:ell_r_gen_b}\\
		&=\frac{\sqrt{r}}{1-\sqrt{r}}\sum_{n=1}^{\infty}\frac{(1-r)^n}{2n}\langle |x_n|\rangle. \label{eq:ell_r_gen}
	\end{align}
\end{subequations}
We remark that both of these equations are valid for any symmetric and continuous jump distribution, not necessarily heavy-tailed, with finite first moment. The first gives $\ell_0(r)$ in terms of $\hat{\lambda}(k)$ only, while the second allows for an analytic computation once $\langle |x_n|\rangle$ is known for each $n$. For instance, in the case of Lévy stable processes, for $1<\mu\leq2$ we have  $\langle|x_n|\rangle=2a\Gamma(1-1/\mu)n^{1/\mu}/\pi$, where $\Gamma(z)$ is the Euler Gamma function. One can then rewrite the series in Eq. \eqref{eq:ell_r_gen} in terms of the polylogarithm $\mathrm{Li}_{1-1/\mu}(1-r)$ \cite{NIST}, and obtain an analytic expression for $\ell_0(r)$, which can be compared with simulations, see Fig. \ref{fig:r_min}. Moreover, in this case  Eq.\eqref{eq:Cond} simplifies to $2^{1/\mu}<\frac32$, which is solved for $\mu>\mu_c=\ln(2)/\ln(3/2)\approx1.7095$. When $\mu>\mu_c$, one can show that $\ell_0(r)$ has indeed a unique minimum attained for $r^*<1$. For $\mu$ below the critical value instead, $\ell_0(r)$ is a decreasing function of $r$, hence $r^*=1$. Thus, we find explicitly that there exists a critical value $\mu_c$ defining a transition to a phase where, by choosing optimally the resetting probability, we obtain an average leapover that is smaller than the average length of a single jump. The whole situation is depicted in Fig. \ref{fig:r_min}, where we plot the  numerically computed $r^*$ versus $\mu$.

We now want to determine explicitly how $\mu_c$ changes by increasing $b$, at least for small $b$. This can be done by approximating the function $\Psi_\mu(b)$ for small $b$ to get from Eq. \eqref{eq:Cond} an explicit condition for the critical exponent $\mu_c(b)$. We linearize $\Psi_\mu(b)\approx\Psi_\mu(0)+\Psi'_\mu(0)b$ and then invert Eq. \eqref{eq:Cond} to write $\mathcal{K}(\mu)=b$, where $\mathcal{K}(\mu)=(\langle|x_2|\rangle/\langle|x_1|\rangle-\Psi_\mu(0))/\Psi'_\mu(0)$. By linearizing then $\mathcal{K}(\mu)$ around $\mu_c$, we finally obtain $\mu_c(b)\approx b/\mathcal{K}'(\mu_c)+\mu_c$. Again, everything can be computed explicitly for Lévy stable processes, and following the procedure we just described we find
\begin{subequations}
	\begin{equation}
		\mu_c(b)\approx \frac{Cb}{a}+\frac{\ln(2)}{\ln(3/2)}\label{eq:mu_c_linear}
	\end{equation}
	\begin{equation}
		C=\frac{\ln(4)}{9\pi\ln^2(3/2)}\Gamma\left(\frac{\ln(3)}{\ln(2)}\right)\approx0.2659973\dots
	\end{equation}
\end{subequations}
The agreement between this linear approximation and the values obtained numerically from the exact condition of Eq. \eqref{eq:Cond} is shown in Fig. \ref{fig:muc_b}.
By following the same idea, we compute an explicit approximation of $\ell_b(r)$ for small $b$. To do this, we need the small-$b$ expansion of $\mathcal{Q}_0(k,z;b)$, which can be obtained from the large-$w$ expansion of its Laplace transform on the rhs of Eq. \eqref{eq:Q_0_b>0}. For continuous and symmetric jump distributions, in the limit $w\to\infty$ the $w$-dependent term at the rhs tends to the Laplace transform of $\sqrt{1-z}\LT{Q}_0(z;b)$, hence we only need a small-$b$ approximation of the generating function. This was calculated in \cite{KusMajSabSch-2014} in the context of search time optimization of Lévy flights. It reads $\sqrt{1-z}\LT{Q}_0(z;b)\approx1+bI(z)+O(b^2)$, where $I(z)$ is an integral that depends on $\hat{\lambda}(k)$. It is then possible for us to write an explicit approximated expression for $\ell_b(r)$, which, by keeping only terms $O(b)$, reads
\begin{equation}\label{eq:ell_b_approx}
	\ell_b(r)\approx\ell_0(r)+b\left(\frac{I(1-r)\ell_0(r)}{1-\sqrt{r}}-1\right).
\end{equation}
In the specific case of Lévy stable processes we can write explicitly $I(z)$ in terms of the polylogarithm $\mathrm{Li}_{1+1/\mu}(z)$, hence we get an analytical curve that we can compare with simulations. The agreement of Eq. \eqref{eq:ell_b_approx} with data is shown in Fig. \ref{fig:muc_b}. We find that the approximation reflects well the behavior of the numerical data for small $b$, although it tends to underestimate them, more so as $b$ increases.

To summarize, we have shown that stochastic resetting is an effective mechanism to control the leapover lengths of one-dimensional Lévy flights performing a myopic search. We have demonstrated that, in contrast to the scenario without resetting, a finite average leapover length is attained and that, moreover, in some cases there exists an optimal resetting probability strictly smaller than one that minimizes the average leapover distance. Consequently, for probabilities close to this optimal value, the average leapover distance is smaller than that obtained by a single jump. A sufficient condition for this is given by Eq. \eqref{eq:Cond}. For Lévy stable processes, we have illustrated our findings in the phase diagram of Fig. \ref{fig:muc_b}. This diagram also shows an interesting phase in which nontrivial optimization is possible even if Eq. \eqref{eq:Cond} is not verified. This work, in conjunction with other research in the literature, opens the way for the study of search strategies aimed at optimizing the coupling of spatial and temporal variables, such as first-passage times and leapover lengths. To expand the generality of our results, it would be beneficial in the future to investigate the phase diagram more thoroughly and for different types of jump distribution, as well as to conduct a rigorous study of the nontrivial phase violating Eq. \eqref{eq:Cond}, which in the present work is only investigated numerically.


%

\ifsupp


\section*{Supplementary material}

# Supplementary Material: "Optimizing Leapover Lengths of Lévy Flights with Resetting"

Mattia Radice

*Max Planck Institute for the Physics of Complex Systems, 01187 Dresden, Germany*

Giampaolo Cristadoro

*Dipartimento di Matematica e Applicazioni, Università degli Studi Milano-Bicocca, Milan, Italy*

## MAIN DEFINITIONS

We first define the quantities that will be used hereafter. We take $x_0 = 0$, set $b \geq 0$ and call $\tau$ the first-passage time of the random walk in $(b, \infty)$. Then $x_\tau$ is the first-passage position in $(b, \infty)$ and $l = x_\tau - b$ is the leapover distance. In absence of resetting, we denote with $f_0(x, n; b)$ the joint probability of leaving the set $(-\infty, b]$ for the first time at step $n$ and landing at position $x > b$, while $q_0(x, n; b)$ is the probability of occupying position $x$ at step $n$ without ever having left $(-\infty, b]$ up to time $n$. Consequently, the survival probability in $n$ steps is

$$Q_0(n; b) = \Pr(\tau > n) = \int_{-\infty}^{b} q_0(x, n; b) dx. \tag{S1}$$

The subfix 0 indicates "absence of resetting", while in presence of resetting the corresponding quantities are denoted $f_r(x, n; b)$ and $q_r(x, n; b)$. However, we simply write $p(x, n)$ for the PDF of the position of the unrestricted reset-free random walk after $n$ steps.

We remark that, according to their definitions, $f_0(x, n; b)$ and $q_0(x, n; b)$ vanish for $x \leq b$ and $x > b$ respectively. In the case $b = 0$, we drop the dependence on $b$ in the notation and simply write, e.g., $f_0(x, n)$ instead of $f_0(x, n; 0)$.

## MAIN FORMULAE

We present here for reference the main quantities that were not displayed in the letter. They will be derived later on in this Supplemental Material. For symmetric and continuous jump distributions with characteristic function $\hat{\lambda}(k) \sim 1 - |ak|^\mu + o(|k|^\mu)$ for small $k$, with $1 < \mu \leq 2$, the average leapover distance in presence of resetting is

$$\ell_b(r) = \frac{r\widetilde{Q}_0(1-r; b) \sum_{n \geq 1} \frac{(1-r)^n}{2n} \langle |x_n| \rangle - h_b(1-r)}{1 - r\widetilde{Q}_0(1-r; b)} - b, \tag{S2}$$

where $\widetilde{Q}_0(z; b) = \sum_{n=0}^{\infty} Q_0(n; b) z^n$ is the survival probability generating function, $\langle |x_n| \rangle = \int_{-\infty}^{+\infty} |x| p(x, n) dx$ is the absolute first moment of the reset-free random walk, and $h_b(z)$ is

$$h_b(z) = -\frac{1}{2\pi} \int_0^b x \widetilde{Q}_0(z; b-x) \int_{-\infty}^{+\infty} \cos(sx) \ln\left(1 - z\hat{\lambda}(s)\right) ds dx = \sum_{n=1}^{\infty} \frac{z^n}{n} \int_0^b x \widetilde{Q}_0(z; b-x) p(x, n) dx. \tag{S3}$$

As mentioned in the letter, $\ell_b(r)$ has a positive derivative for $r \to 1^-$ if $\langle |x_2| \rangle / \langle |x_1| \rangle < \Psi_\mu(b)$, where

$$\Psi_\mu(b) = 2 - 2\frac{Q_0(2; b) - Q_0(1; b)^2}{1 - Q_0(1; b)} + \frac{\int_0^b xp(x, 2) dx}{\int_0^\infty x\lambda(x) dx} + 2\frac{\int_0^b x\lambda(x)[Q_0(1; b-x) - \frac{1-Q_0(2;b)}{1-Q_0(1;b)}] dx}{\int_0^\infty x\lambda(x) dx}. \tag{S4}$$

## DERIVATION OF EQ.(3) IN THE MAIN TEXT

We start from Eq. (2) in the main text

$$f_r(x, n; b) = r \sum_{m=0}^{n-1} (1-r)^m Q_0(m; b) f_r(x, n-m; b) + (1-r)^n f_0(x, n; b). \quad n \geq 1, \tag{S5}$$

To explain this, note that according to Eq.(1) in the main text, we are using a resetting protocol where first the position is reset, and then the jump is performed. If no resetting occurs before the first-passage time, then simply $f_r(x, n; b) = (1 - r)^n f_0(x, n; b)$. Otherwise, we can assume that the first resetting occurs before the $(m + 1)$-th step, with $m \geq 0$. Then the walk survived for $m$ steps and then restarted, having $m - n$ steps to cross $b$.

By introducing the double transform $\mathcal{G}(k, z) = \sum_{n=0}^{\infty} z^n \int_{-\infty}^{+\infty} e^{ikx} g(x, n) dx$, Eq. (S5) becomes

$$\mathcal{F}_r(k, z; b) = \frac{\mathcal{F}_0(k, z_r; b)}{1 - r \mathcal{Q}_0(0, z_r; b)}, \quad z_r = (1 - r) z, \tag{S6}$$

where we used $\widetilde{Q}_0(z; b) = \mathcal{Q}_0(0, z; b)$. Then we use a known relation between the two transforms $\mathcal{F}_0(k, z; b)$ and $\mathcal{Q}_0(k, z; b)$, see for instance [1], which reads

$$\frac{1}{1 - z \hat{\lambda}(k)} = \frac{\mathcal{Q}_0(k, z; b)}{1 - \mathcal{F}_0(k, z; b)}. \tag{S7}$$

By isolating $\mathcal{F}_0(k, z; b)$ and substituting into Eq. (S6), we obtain $\mathcal{F}_r(k, z; b)$ in terms of $\mathcal{Q}_0(k, z_r; b)$:

$$\mathcal{F}_r(k, z; b) = \frac{1 - [1 - (1 - r) z \hat{\lambda}(k)] \mathcal{Q}_0(k, z_r; b)}{1 - r \mathcal{Q}_0(0, z_r; b)}. \tag{S8}$$

We then get Eq. (3) by noting that $\wp_r(l; b) = \sum_{n=1}^{\infty} f_r(l + b, n; b) = \widetilde{f}_r(l + b, 1; b)$, hence by taking the Fourier transform we get

$$\widehat{\wp}_r(k; b) = \int_0^{\infty} e^{ikl} \widetilde{f}_r(l + b, 1; b) dl = e^{-ikb} \int_b^{\infty} e^{ikx} \widetilde{f}_r(x, 1; b) dx = e^{-ikb} \mathcal{F}_r(k, 1; b). \tag{S9}$$

## DERIVATION OF EQ. (4) IN THE MAIN TEXT

In this section it is implied that we are considering only the reset-free walk, hence $x_n = \sum_{n \geq 0} \eta_n$, with $\eta_n$ independent and identically distributed random variables drawn from $\lambda(\eta)$. We denote with $b_n$ the maximum of the random walk up to step $n$; we call $x_n^+$ the maximum between 0 and $x_n$, with characteristic function $u_n(k) = \mathbb{E} \exp(ikx_n^+)$; similarly, we call $x_n^-$ the maximum between 0 and $-x_n$, with characteristic function $v_n(k) = \mathbb{E} \exp(ikx_n^-)$. Furthermore, we introduce

$$\widehat{g}(\alpha, \beta, n) = \mathbb{E} \left( e^{i\alpha b_n + i\beta(b_n - x_n)} \right). \tag{S10}$$

We will use the following theorem (Spitzer, Theorem 6.1 in [2]):

$$\sum_{n=0}^{\infty} z^n \widehat{g}(\alpha, \beta, n) = \exp \left[ \sum_{n=1}^{\infty} \frac{z^n}{n} \left( u_n(\alpha) + v_n(\beta) - 1 \right) \right]. \tag{S11}$$

It is immediately seen that $\widehat{g}(\alpha, \beta, n)$ is related to the joint characteristic function of $b_n$ and $x_n$ by

$$\mathbb{E} \left( e^{itb_n + ikx_n} \right) = \mathbb{E} \left( e^{i(k+t)b_n - ik(b_n - x_n)} \right) = \widehat{g}(k + t, -k, n), \tag{S12}$$

and we note that $x_n^+ = 0 (x_n^- = 0)$ whenever $x_n \leq 0 (x_n \geq 0)$ and $x_n^+ = x_n (x_n^- = -x_n)$ otherwise. Then

$$u_n(k) = \Pr(x_n \leq 0) + \int_0^{\infty} e^{ikx} p(x, n) dx, \tag{S13}$$

$$v_n(k) = \Pr(x_n \geq 0) + \int_{-\infty}^0 e^{-ikx} p(x, n) dx, \tag{S14}$$

where $p(x, n)$ is the PDF of the position of the random walk in $n$ steps. Define now

$$\phi_+(k, z) = \sum_{n=1}^{\infty} \frac{z^n}{n} \int_0^{\infty} e^{ikx} p(x, n) dx \tag{S15}$$

$$\phi_-(k, z) = \sum_{n=1}^{\infty} \frac{z^n}{n} \int_{-\infty}^0 e^{ikx} p(x, n) dx, \tag{S16}$$

so that, from Eq. (S11) and Eq. (S12) we can introduce $\mathcal{G}(k,t,z) = \sum_{n\geq 0} z^n \mathbb{E}\left(e^{itb_n + ikx_n}\right)$ and get

$$\mathcal{G}(k,t,z) = \sum_{n=0}^{\infty} z^n \hat{g}(k+t,-k,n) = e^{\phi_+(k+t,z)} e^{\phi_-(k,z)}. \tag{S17}$$

We observe that $\phi_\pm(\xi,z)$ are analytic and bounded functions for $\pm\text{Im}(\xi) > 0$, respectively, hence we can extend $\mathcal{G}(k,t,z)$ to $\text{Im}(t) > 0$. In particular, note that for $t = iw$, with $w > 0$, we have

$$\mathcal{G}(k,iw,z) = \sum_{n=0}^{\infty} z^n \mathbb{E}\left(e^{-wb_n + ikx_n}\right) = \sum_{n=0}^{\infty} z^n \int_0^\infty e^{-wb} \int_{-\infty}^{+\infty} e^{ikx} \Pr(x_n = x, b_n = b) dx db$$

$$= w \sum_{n=0}^{\infty} z^n \int_{-\infty}^{+\infty} \int_0^\infty e^{ikx} e^{-wb} \Pr(x_n = x, b_n \leq b) dx db$$

$$= w \int_0^\infty e^{-wb} \mathcal{Q}_0(k,z;b) db, \tag{S18}$$

whence it follows

$$w \int_0^\infty e^{-wb} \mathcal{Q}_0(k,z;b) db = e^{\phi_+(k+iw,z)} e^{\phi_-(k,z)}. \tag{S19}$$

Equation (4) in the main text follows from the identity $e^{\phi_-(k,z)} = \mathcal{Q}_0(k,z)$, which is a classical result [2]. We remark that here we have not assumed that $\lambda(\eta)$ is symmetric, hence Eq. (4) is also valid for asymmetric jump distributions.

## DERIVATION OF EQ. (5) IN THE MAIN TEXT

The functions $\phi_\pm(k,z)$ can be rewritten in terms of $\hat{\lambda}(k)$ instead of $p(x,n)$. For instance, since $\phi_+(\xi,z)$ is analytic for $\text{Im}(\xi) > 0$, for any complex $\xi$ with $\text{Im}(\xi) > 0$ we can use $p(x,n) = \frac{1}{2\pi} \int_{-\infty}^{+\infty} e^{-ikx} \hat{\lambda}(k)^n dk$ to write

$$\phi_+(\xi,z) = \frac{1}{2\pi} \sum_{n=1}^{\infty} \frac{z^n}{n} \int_0^\infty e^{i\xi x} \int_{-\infty}^{+\infty} e^{-isx} \hat{\lambda}(s)^n ds dx = \frac{1}{2\pi i} \int_{-\infty}^{+\infty} \frac{ds}{\xi - s} \ln\left(1 - z\hat{\lambda}(s)\right). \tag{S20}$$

For the same reason, for any $\xi$ with $\text{Im}(\xi) < 0$ one has

$$\phi_-(\xi,z) = -\frac{1}{2\pi i} \int_{-\infty}^{+\infty} \frac{ds}{\xi - s} \ln\left(1 - z\hat{\lambda}(s)\right). \tag{S21}$$

We point out that by using the analytical properties of $\phi_\pm(k,z)$ in combination with Eq. (4), we can recover some known results [2–8]. We also remark that $\phi_\pm(k,z)$ are related to the Wiener-Hopf factors of $1/(1 - z\hat{\lambda}(k))$, which is the double transform of $p(x,n)$. That is, we have the identity

$$\sum_{n=0}^{\infty} z^n \int_{-\infty}^{+\infty} e^{ikx} p(x,n) dx = \frac{1}{1 - z\hat{\lambda}(k)} = e^{\phi_+(k,z)} e^{\phi_-(k,z)}, \tag{S22}$$

hence $\phi_\pm(\xi,z)$ are easy to obtain whenever this factorization can be written explicitly. For example, in the case of the symmetric exponential distribution, which is one of the few cases that is explicitly solvable [9], we have $\hat{\lambda}(k) = 1/(1 + a^2 k^2)$, hence

$$\frac{1}{1 - z\hat{\lambda}(k)} = \frac{a^2 k^2 + 1}{a^2 k^2 + 1 - z} = \frac{(ak + i)(ak - i)}{(ak + i\sqrt{1-z})(ak - i\sqrt{1-z})}, \tag{S23}$$

thus

$$\phi_\pm(\xi,z) = \ln\left(\frac{a\xi \pm i}{a\xi \pm i\sqrt{1-z}}\right). \tag{S24}$$

## EXPANSION OF THE LEAPOVER DISTRIBUTION

To derive the small-$k$ expansion of $\widehat{\wp}_r(k; b)$, we first need that of $\mathcal{Q}_0(k, z; b)$. From Eq. (4) in the main text, we see that we must compute the expansion of a $b$-independent part, given by $\mathcal{Q}_0(k, z)$ and a $b$-dependent part. In the following we analyze the two parts separately. We only consider symmetric jump distributions with characteristic function $\hat{\lambda}(k) \sim 1 - |ak|^\mu$ for small $k$, with $0 < \mu < 2$.

### Expansion of $\mathcal{Q}_0(k, z)$

To find the small-$k$ expansion of $\mathcal{Q}_0(k, z)$ we can use [10]

$$\mathcal{Q}_0(k, z) = \frac{e^{-i\Phi(k,z)}}{\sqrt{1 - z\hat{\lambda}(k)}}, \tag{S25}$$

where $\Phi(k, z)$ is the *phase* function defined by

$$\Phi(k, z) = \frac{1}{2\pi} \text{P.V.} \int_{-\infty}^{+\infty} \frac{ds}{k - s} \ln\left(1 - z\hat{\lambda}(s)\right). \tag{S26}$$

Here P.V. denotes the Cauchy principal value of the integral. Equation (S25) was obtained by Zumofen and Klafter [10] with standard Fourier methods. In our setting, it is sufficient to consider the identity $\mathcal{Q}_0(\xi, z) = e^{\phi_-(\xi, z)}$ for $\text{Im}(\xi) < 0$ and then take the limit $\text{Im}(\xi) \to 0^-$, which is captured by the Plemelj formulae [11]. We first observe that for symmetric $\hat{\lambda}(k)$, the phase is an odd function of $k$, hence to leading order it does not contribute to the small-$k$ expansion of the real part, which is then

$$\text{Re}(\mathcal{Q}_0(k, z)) \sim \frac{1}{\sqrt{1 - z}} \left[1 - \frac{z|ak|^\mu}{2(1 - z)}\right]. \tag{S27}$$

For the imaginary part we need instead the expansion of the phase for fixed $0 \le z < 1$. We use $\ln(1 - z\hat{\lambda}(k)) = \sum_{n \ge 1} z^n \hat{\lambda}(k)^n / n$ and the symmetry of $\hat{\lambda}(k)$ to rewrite

$$\Phi(k, z) = \frac{k}{\pi} \sum_{n=1}^{\infty} \frac{z^n}{n} \text{P.V.} \int_0^\infty \frac{\hat{\lambda}(s)^n}{k^2 - s^2} ds. \tag{S28}$$

For the purpose of the letter, it is relevant only the restricted range $1 < \mu \le 2$. We sum and subtract 1 in the integral at the rhs and note that the principal value of $1/(k^2 - s^2)$ vanishes. Then we divide both sides for $k$ and take the limit $k \to 0$, obtaining

$$\lim_{k \to 0} \frac{\Phi(k, z)}{k} = \lim_{k \to 0} \frac{1}{\pi} \sum_{n=1}^{\infty} \frac{z^n}{n} \text{P.V.} \int_0^\infty \frac{\hat{\lambda}(s)^n - 1}{k^2 - s^2} ds = \frac{1}{\pi} \sum_{n=1}^{\infty} \frac{z^n}{n} \int_0^\infty \frac{1 - \hat{\lambda}(s)^n}{s^2} ds = \sum_{n=1}^{\infty} \frac{z^n}{2n} \langle |x_n| \rangle. \tag{S29}$$

The last equality holds because

$$\frac{1}{\pi} \int_0^\infty ds \int_{-\infty}^{+\infty} dx \frac{1 - \cos(sx)}{s^2} p(x, n) = \frac{1}{\pi} \int_0^\infty \frac{1 - \hat{\lambda}(s)^n}{s^2} ds = \frac{1}{2} \int_{-\infty}^{+\infty} |x| p(x, n) dx, \tag{S30}$$

since one is allowed to interchange the order of integration in the first integral. It follows that $\Phi(k) \sim k \sum_{n \ge 1} \frac{z^n}{2n} \langle |x_n| \rangle$ and then

$$\text{Im}(\mathcal{Q}_0(k, z)) \sim -\frac{k}{\sqrt{1 - z}} \sum_{n=1}^{\infty} \frac{z^n}{2n} \langle |x_n| \rangle. \tag{S31}$$

## Expansion of the $b$-dependent part

To simplify the notation, let us call $\mathcal{H}(k, b, z)$ the $b$-dependent part, defined by

$$\int_0^\infty e^{-wb}\mathcal{H}(k, b, z)db = \frac{1}{w}e^{\phi_+(k+iw,z)}. \tag{S32}$$

Since $\phi_+(\xi, z)$ is analytic for $\mathrm{Im}(\xi) > 0$, when $w > 0$ both sides posses an analytic expansion in powers of $k$ around $k = 0$. Hence we can rewrite the rhs as

$$\frac{1}{w}\exp\left(\sum_{n=0}^\infty \frac{\phi_+^{(n)}(iw, z)}{n!}k^n\right), \tag{S33}$$

where

$$\frac{\phi_+^{(n)}(\xi, z)}{n!} = \frac{(-1)^n}{2\pi i}\int_{-\infty}^{+\infty} \frac{ds}{(\xi - s)^{n+1}}\ln\left(1 - z\hat{\lambda}(s)\right), \quad \mathrm{Im}(\xi) > 0. \tag{S34}$$

Expanding up to terms of order $k^2$ we have

$$\mathrm{Re}\left(\frac{1}{w}e^{\phi_+(k+iw,z)}\right) \sim \frac{e^{\phi_+(iw,z)}}{w} + O(k^2) \tag{S35}$$

$$\mathrm{Im}\left(\frac{1}{w}e^{\phi_+(k+iw,z)}\right) \sim k\frac{e^{\phi_+(iw,z)}}{2\pi w}\int_{-\infty}^{+\infty} \frac{s^2 - w^2}{(s^2 + w^2)^2}\ln\left(1 - z\hat{\lambda}(s)\right)ds + O(k^2). \tag{S36}$$

Note that the rhs of Eq. (S35) can be rewritten as

$$\frac{e^{\phi_+(iw,z)}}{w} = \frac{e^{\phi_-(-iw,z)}}{w} = \frac{\mathcal{Q}_0(-iw, z)}{w}, \tag{S37}$$

where the chain of identities follows from symmetry. Now note that

$$\frac{\mathcal{Q}_0(-iw, z)}{w} = \frac{1}{w}\exp\left(\frac{1}{2\pi i}\int_{-\infty}^{+\infty} \frac{ds}{iw + s}\ln\left(1 - z\hat{\lambda}(s)\right)\right) = \frac{1}{w}\exp\left(\frac{w}{2\pi}\int_{-\infty}^{+\infty} \frac{ds}{w^2 + s^2}\ln\left(1 - z\hat{\lambda}(s)\right)\right), \tag{S38}$$

where, from the Pollaczek-Spitzer formula [2, 4, 6], the last term is the Laplace transform of $\sqrt{1 - z}\widetilde{Q}_0(z; b)$. The Laplace inversion thus yields

$$\mathrm{Re}(\mathcal{H}(k, b, z)) \sim \sqrt{1 - z}\widetilde{Q}_0(z; b) + O(k^2). \tag{S39}$$

The imaginary part instead can be formally written as

$$\mathrm{Im}\left(\mathcal{H}(k, b, z)\right) \sim \frac{k}{2\pi i}\int_{\gamma - i\infty}^{\gamma + i\infty} \frac{e^{wb}}{2\pi w}e^{\phi_+(iw,z)}\int_{-\infty}^{+\infty} \frac{s^2 - w^2}{(w^2 + s^2)^2}\ln\left(1 - z\hat{\lambda}(s)\right)dsdw + O(k^2). \tag{S40}$$

To perform the inversion, we first integrate in $w$. We have the product of two functions: the first is again the Laplace transform of $\sqrt{1 - z}\widetilde{Q}(z; b)$, while the second is the Laplace transform of $-b\cos(bs)$. Therefore, the inversion yields the convolution, and we eventually obtain

$$\mathrm{Im}\left(\mathcal{H}(k, b, z)\right) \sim k\sqrt{1 - z}h_b(z) + O(k^2), \tag{S41}$$

where $h_b(z)$ is given by Eq. (S3).

## Expansion of $\widehat{\varphi}_r(k; b)$

We are now ready to compute the small-$k$ expansion of $\widehat{\varphi}_r(k; b)$. First, let us collect the previous results by writing the expansions of the real and imaginary parts of $\mathcal{Q}_0(k, z; b)$, which to leading order read

$$\mathrm{Re}(\mathcal{Q}_0(k, z; b)) \sim \widetilde{Q}_0(z; b)\left(1 - \frac{z|ak|^\mu}{2(1 - z)}\right) \tag{S42}$$

$$\mathrm{Im}(\mathcal{Q}_0(k, z; b)) \sim k\left(h_b(z) - \widetilde{Q}_0(z; b)\sum_{n=1}^\infty \frac{z^n}{2n}\langle|x_n|\rangle\right). \tag{S43}$$

By taking the small-$k$ expansion of Eq. (4) in the main text we thus get

$$\mathrm{Re}(\widehat{\wp}_r(k;b)) \sim 1 - A_r(b)|ak|^\mu = 1 - \frac{1}{2}\frac{(1-r)\widetilde{Q}_0(1-r;b)}{1-r\widetilde{Q}_0(1-r;b)}|ak|^\mu \tag{S44}$$

$$\mathrm{Im}(\widehat{\wp}_r(k;b)) \sim k\ell_b(r) = k\left(\frac{r\widetilde{Q}_0(1-r;b)\sum_{n\geq 1}\frac{z^n}{2n}\langle|x_n|\rangle - h_b(1-r)}{1-r\widetilde{Q}_0(1-r;b)} - b\right), \tag{S45}$$

as announced in the letter. We can easily check what these expressions yield the correct result for $r=1$. Indeed, In this regime we have to recover the leapover distribution of the single jump, which is

$$\widehat{\wp}_1(k;b) = e^{-ikb}\frac{\int_b^\infty e^{ikx}\lambda(x)dx}{\int_b^\infty \lambda(x)dx}, \tag{S46}$$

hence if $\lambda(k) \sim 1 - |ak|^\mu$, with $1 < \mu < 2$, we must have

$$\mathrm{Re}(\widehat{\wp}_1(k;b)) \sim 1 - \frac{|ak|^\mu}{2\int_b^\infty \lambda(x)dx} \tag{S47}$$

$$\mathrm{Im}(\widehat{\wp}_1(k;b)) \sim k\left(\frac{\int_b^\infty x\lambda(x)dx}{\int_b^\infty \lambda(x)dx} - b\right). \tag{S48}$$

Both are recovered from Eqs. (S44) and (S45) by using that for small $z$:

$$\widetilde{Q}(z;b) = 1 + z\int_{-\infty}^b \lambda(x)dx + O(z^2). \tag{S49}$$

## PROOF OF EQ. (6) IN THE MAIN TEXT

To obtain the condition for a positive derivative of $\ell_b(r)$ when $r \to 1^-$, we simply expand Eq. (S2) in powers of $1-r$. This yields

$$\ell_b(r) \sim \frac{\int_b^\infty x\lambda(x)dx}{\int_b^\infty \lambda(x)dx} - b + \frac{(1-r)}{\int_b^\infty \lambda(x)dx}\left[\frac{\langle|x_2|\rangle}{4} - \frac{\langle|x_1|\rangle}{2}\left(1 - \frac{Q_0(2;b)-Q_0(1;b)^2}{1-Q_0(1;b)}\right) - \frac{1}{2}\int_0^b xp(x,2)dx\right.$$
$$\left. - \int_0^b x\lambda(x)\left(Q_0(1;b-x) - \frac{1-Q_0(2;b)}{1-Q_0(1;b)}\right)dx\right] + O((1-r)^2). \tag{S50}$$

The desired condition is then obtained by imposing that the term inside the square brackets is negative. One obtains Eq. (6), with $\Psi_\mu(b)$ given by Eq. (S4).

## EXPLICIT EXPRESSIONS FOR $b=0$

For $b=0$, we can use the Sparre-Andersen universality [12] to write explicitly the generating function $\widetilde{Q}_0(z;b=0) = 1/\sqrt{1-z}$. This is true for any continuous and symmetric jump distribution. Then from Eq. (S2) we get

$$\ell_0(r) = \frac{\sqrt{r}}{1-\sqrt{r}}\sum_{n=1}^\infty \frac{(1-r)^n}{2n}\langle|x_n|\rangle. \tag{S51}$$

By recalling that

$$\frac{1}{2}\langle|x_n|\rangle = \frac{1}{\pi}\int_0^\infty \frac{1-\hat\lambda(s)^n}{s^2}ds, \tag{S52}$$

see Eq. (S29), we can rewrite

$$\ell_0(r) = \frac{\sqrt{r}}{\pi(1-\sqrt{r})}\sum_{n=1}^\infty \frac{(1-r)^n}{n}\int_0^\infty \frac{1-\hat\lambda(s)^n}{s^2}ds = \frac{\sqrt{r}}{\pi(1-\sqrt{r})}\int_0^\infty \frac{ds}{s^2}\ln\left(\frac{1-(1-r)\hat\lambda(s)}{r}\right). \tag{S53}$$

For Lévy stable processes, with $\hat{\lambda}(k) = e^{-|ak|^\mu}$, in the range $1 < \mu \leq 2$ we have

$$\frac{1}{2}\langle|x_n|\rangle = \frac{1}{\pi}\int_0^\infty \frac{1-e^{-n(as)^\mu}}{s^2}ds = \frac{a}{\pi}\Gamma\left(1-\frac{1}{\mu}\right)n^{1/\mu}, \tag{S54}$$

hence

$$\ell_0(r) = \frac{a}{\pi}\Gamma\left(1-\frac{1}{\mu}\right)\frac{\sqrt{r}}{1-\sqrt{r}}\sum_{n=1}^\infty \frac{(1-r)^n}{n^{1-1/\mu}} = \frac{a}{\pi}\Gamma\left(1-\frac{1}{\mu}\right)\frac{\sqrt{r}}{1-\sqrt{r}}\text{Li}_{1-1/\mu}(1-r). \tag{S55}$$

One can verify from Eq. (S51) that when $r \to 1^-$

$$\ell_0(r) \sim \langle|x_1|\rangle + \left(\frac{1}{4}\langle|x_2|\rangle - \frac{3}{8}\langle|x_1|\rangle\right)(1-r) + O((1-r)^2), \tag{S56}$$

therefore the derivative is positive if

$$\frac{\langle|x_2|\rangle}{\langle|x_1|\rangle} < \frac{3}{2}. \tag{S57}$$

The same condition can be obtained from Eq. (S4) by setting $b = 0$ and using $\widetilde{Q}_0(z;0) = 1/\sqrt{1-z}$ to obtain $Q_0(1;0) = \frac{1}{2}$ and $Q_0(2;0) = \frac{3}{8}$. For Lévy stable processes one can write explicitly

$$2^{1/\mu} < \frac{3}{2}, \tag{S58}$$

which is solved for $\mu > \mu_c = \ln 2/\ln(3/2)$.

## APPROXIMATION FOR SMALL $b$

Starting from Eq. (4) in the main text, we seek a small-$b$ approximation of $\mathcal{Q}_0(k,z;b)$ by looking at the large-$w$ expansion of the rhs and then performing the Laplace inversion. If we assume that the process is symmetric, we can use $\phi_+(\xi,z) = \phi_-(-\xi,z)$ to see that in the limit $w \to \infty$, with $w \gg k$:

$$\frac{e^{\phi_+(k+iw,z)}}{w} \sim \frac{e^{\phi_-(-iw,z)}}{w} = \frac{\mathcal{Q}_0(-iw,z)}{w}, \tag{S59}$$

hence, as a consequence of the Pollaczek-Spitzer formula, upon inversion we can write for small $b$

$$\mathcal{Q}_0(k,z;b) \approx \sqrt{1-z}\,\mathcal{Q}_0(k,z)\widetilde{Q}_0(z;b). \tag{S60}$$

The small-$b$ approximation of the lhs thus follows from that of $\widetilde{Q}_0(z;b)$, which was computed in [13]. One gets

$$\mathcal{Q}_0(k,z;b) \approx \mathcal{Q}_0(k,z)\left(1 + bI(z)\right), \tag{S61}$$

with

$$I(z) = -\frac{1}{2\pi}\int_{-\infty}^{+\infty}\ln\left(1 - z\hat{\lambda}(s)\right)ds. \tag{S62}$$

In general, by recalling that

$$\phi_+(\xi,z) = \frac{1}{2\pi i}\int_{-\infty}^{+\infty}\frac{ds}{\xi-s}\ln\left(1-z\hat{\lambda}(s)\right), \tag{S63}$$

we can use the fact that when $|\xi| \to \infty$ off the integration contour, then $\phi_+(\xi,z) \sim c(z)/\xi$, with $c(z) = \frac{1}{2\pi i}\int_{-\infty}^{+\infty}\ln\left(1 - z\hat{\lambda}(s)\right)ds$, see [11]. Hence in the limit $w \to \infty$, with $w \gg k$, we obtain

$$\frac{e^{\phi_+(k+iw,z)}}{w} \sim \frac{e^{c(z)/iw}}{w} \sim \frac{1}{w} + \frac{I(z)}{w^2}, \tag{S64}$$

and Eq. (S61) is obtained upon inversion.

**Small-$b$ approximation of Eq. (6)**

To find an approximation for the condition of Eq. (6), it is sufficient to find the small-$b$ expansion of $\Psi_\mu(b)$ given by Eq. (S4). We are interested in terms $O(b)$, which only come from

$$2 - 2\frac{Q_0(2;b) - Q_0(1;b)^2}{1 - Q_0(1;b)}, \tag{S65}$$

hence we need the small-$b$ approximation of the survival probabilities $Q_0(1;b)$ and $Q_0(2;b)$. From the Pollaczek-Spitzer formula we have exactly

$$Q_0(1;b) = \frac{1}{2} + \frac{1}{\pi}\int_0^\infty \frac{ds}{s}\sin(bs)\hat{\lambda}(s) \tag{S66}$$

$$Q_0(2;b) = \frac{3}{8} + \frac{1}{2\pi}\int_0^\infty \frac{ds}{s}\sin(bs)\left(\hat{\lambda}(s) + \hat{\lambda}(s)^2\right) + \int_{\gamma-i\infty}^{\gamma+i\infty}\frac{dw}{2\pi i}e^{wb}\left(\frac{w}{2\pi}\int_0^\infty \frac{ds}{w^2+s^2}\hat{\lambda}(s)\right)^2. \tag{S67}$$

Note that in the first line we have the survival probability of the first jump, which is equal to $\int_{-\infty}^b \lambda(x)dx = \frac{1}{2} + \int_0^b \lambda(x)dx$. Indeed

$$\begin{aligned}\frac{1}{\pi}\int_0^\infty \frac{ds}{s}\sin(bs)\hat{\lambda}(s) &= \frac{b}{2\pi}\int_{-\infty}^{+\infty}ds\left(\int_{-\infty}^{+\infty}dx e^{isx}\frac{\Theta(|x|\le b)}{2b}\right)\hat{\lambda}(s)\\ &= \frac{b}{2\pi}\int_{-\infty}^{+\infty}dx\left(\int_{-\infty}^{+\infty}ds e^{isx}\hat{\lambda}(s)\right)\frac{\Theta(|x|\le b)}{2b}\\ &= b\int_{-\infty}^{+\infty}dx\lambda(-x)\frac{\Theta(|x|\le b)}{2b} = \frac{1}{2}\int_{-b}^b \lambda(x)dx = \int_0^b dx\lambda(x).\end{aligned} \tag{S68}$$

Hence when $b$ is small, the first line yields $Q_0(1;b) \approx \frac{1}{2} + b\lambda(0)$. In the second line we can neglect the last term, which is $O(b^2)$, to get in a similar way $Q_0(2;b) \approx \frac{3}{8} + \frac{1}{2}\lambda(0) + \frac{1}{2}p(0,2)$. Therefore

$$\Psi_\mu(b) \approx \frac{3}{2} + b\left(\lambda(0) - 2p(0,2)\right). \tag{S69}$$

For Lévy stable processes, we have explicitly

$$p(0,n) = \frac{1}{2\pi}\int_{-\infty}^{+\infty}\hat{\lambda}(s)^n ds = \frac{1}{2\pi}\int_{-\infty}^{+\infty}e^{-n|as|^\mu}ds = \frac{n^{-1/\mu}}{a\pi}\Gamma\left(1+\frac{1}{\mu}\right), \tag{S70}$$

and we recall $\langle|x_n|\rangle = 2a\Gamma(1-1/\mu)n^{1/\mu}/\pi$, hence an explicit approximation of Eq. (6) is

$$2^{1/\mu} < \frac{3}{2} + \frac{b}{a\pi}\Gamma\left(1+\frac{1}{\mu}\right)\left(1-2^{1-1/\mu}\right). \tag{S71}$$

By equating both sides of this inequality we get the equation for the critical exponent $\mu_c(b)$. To write it down explicitly, we isolate $b$ to get

$$b = a\pi\frac{2^{1/\mu} - 3/2}{\Gamma(1+1/\mu)(1-2^{1-1/\mu})} = \mathcal{K}(\mu), \tag{S72}$$

and then expand the rhs around $\mu_c = \ln(2)/\ln(3/2)$. Since $\mathcal{K}(\mu_c) = 0$ by definition, we have

$$b = \mathcal{K}'(\mu_c)(\mu_c(b) - \mu_c), \tag{S73}$$

whence Eqs. (8a)-(8b) follow.

## Small-$b$ approximation of $\ell_b(r)$

By plugging Eq. (S61) into Eq. (3) in the main text and evaluating the small-$k$ expansion of the imaginary part, we find the approximated expression of $\ell_b(r)$ given by Eq. (9) in the main text. For Lévy stable processes we get an analytic expression by computing $I(z)$ explicitly:

$$
\begin{aligned}
I(z) = -\frac{1}{\pi} \int_0^\infty \ln\left(1 - z e^{-(as)^\mu}\right) ds &= \frac{1}{\pi} \sum_{n=1}^\infty \frac{z^n}{n} \int_0^\infty e^{-n(as)^\mu} ds \\
&= \frac{1}{a\pi\mu} \sum_{n=1}^\infty \frac{z^n}{n^{1+1/\mu}} \int_0^\infty e^{-s} s^{1/\mu-1} ds \\
&= \frac{1}{a\pi} \Gamma\left(1 + \frac{1}{\mu}\right) \mathrm{Li}_{1+1/\mu}(z).
\end{aligned}
\tag{S74}
$$

---


\fi

\end{document}